%% file: naacl2021.tex
\pdfoutput=1

\documentclass[11pt]{article}

\usepackage{naacl2021}

\usepackage{times}
\usepackage{latexsym}
\usepackage{tabularx}
\usepackage{comment}
\usepackage{listings}
\usepackage{adjustbox,lipsum}

\lstnewenvironment{queryl}[1][]
 {\lstset{frame=single,escapechar=`,linewidth=\textwidth, #1}}
 {}
 
\usepackage[T1]{fontenc}
\usepackage[utf8]{inputenc}
\usepackage{microtype}
\usepackage{graphicx}
\usepackage{booktabs}
\usepackage{multirow}
\usepackage{amsmath}

\title{ProTIP: Progressive Tool Retrieval Improves Planning}

\author{Raviteja Anantha*, Bortik Bandyopadhyay*, Anirudh Kashi, Sayantan Mahinder, \\
\textbf{Andrew W Hill}, \textbf{Srinivas Chappidi} \\
Apple}

\begin{document}
\maketitle
\begin{abstract}
Large language models (LLMs) are increasingly employed for complex multi-step planning tasks, where the tool retrieval (TR) step is crucial for achieving successful outcomes. Two prevalent approaches for TR are single-step retrieval, which utilizes the complete query, and sequential retrieval using task decomposition (TD), where a full query is segmented into discrete atomic subtasks. While single-step retrieval lacks the flexibility to handle ``inter-tool dependency,"  the TD approach necessitates maintaining ``subtask-tool atomicity alignment," as the toolbox can evolve dynamically. To address these limitations, we introduce the \textbf{Pro}gressive \textbf{T}ool retrieval to \textbf{I}mprove \textbf{P}lanning (ProTIP) framework. ProTIP is a lightweight, contrastive learning-based framework that implicitly performs TD without the explicit requirement of subtask labels, while simultaneously maintaining subtask-tool atomicity. On the ToolBench dataset, ProTIP outperforms the ChatGPT task decomposition-based approach by a remarkable margin, achieving a 24\% improvement in Recall@K=10 for TR and a 41\% enhancement in tool accuracy for plan generation.

\end{abstract}

\def\thefootnote{*}\footnotetext{Equal contributions.}

\input{introduction}

\input{methodology}

\input{results}

\input{related_work}
\input{conclusion}

\bibliography{custom}
\bibliographystyle{acl_natbib}

\include{appendix}

\end{document}

%% file: introduction.tex
\section{Introduction}

Large language models (LLMs)~\cite{brown2020language:20,  chowdhery2022palm, ouyang2022training, zhang2022opt, zeng2022glm, openai2023gpt, touvron2023llama, bubeck2023sparks, thoppilan2022lamda} have witnessed remarkable advancements in recent years, driven by efficient pre-training on massive text datasets and the application of sophisticated algorithms like reinforcement learning from human feedback (RLHF)~\cite{ouyang2022training} to better align these models with human preferences. This progress has unlocked emergent capabilities~\cite{wei2022emergent} in LLMs, which can be further refined to enhance their instruction-following abilities~\cite{alpaca, vicuna2023}. Additionally, LLMs have demonstrated the potential for in-context learning~\cite{brown2020language:20, xie2021explanation, minetal2022rethinking} and chain-of-thought prompting~\cite{wei2022chain, kojima2022large, wang2022self}, expanding their applicability across a wide spectrum of application domains.

Harnessing the power of LLMs as language understanding agent~\cite{shen2023hugginggpt} to tackle complex tasks has emerged as a burgeoning research area. This endeavor presents a challenge due to the inherent complexity of multi-step planning~\cite{huang2022language, ahn2022i, singh2022progprompt}. 
To address this challenge, we employ a flexible planning framework that seamlessly integrates an LLM with an external toolbox containing application specific atomic actions.
The LLM planner bridges the gap between natural language instructions and executable actions by effectively selecting appropriate APIs/tools from a curated list presented in the LLM prompt. These tools are retrieved using specialized techniques from the external toolbox~\cite{schick2023toolformer, qin2023tool, patil2023gorilla, toolbench:23, shen2023hugginggpt}. The terms tool and API are used interchangeably throughout this paper.

Within multi-step planning framework with an external toolbox, the tool retrieval (TR) step plays a crucial role in determining the overall planner's performance.
The TR step can be implemented either as a single-step process utilizing the entire query or as an iterative approach that decomposes the query into individual atomic subtasks~\cite{decomposed-prompting:2022, pet:23}.
The single-step TR approach, however, is unable to handle ``inter-tool dependency" in multi-step execution scenarios. This limitation becomes evident, for instance, when selecting between tool-A and tool-B, where the choice depends on the successful execution of a previously chosen tool. 
In contrast, the TD-based TR approach necessitates maintaining the alignment between the exact subtask in question and the appropriate tool to be used from the employed toolbox version, thus creating a ``subtask-tool atomicity alignment," problem when training the planner. This dependency often requires either frequent fine-tuning of lightweight TD models or the utilization of an LLM, such as ChatGPT~\cite{chatgpt:23}, for TD.
Furthermore, both these approaches operate within the text space, making them susceptible to various issues such as ``out of vocabulary" tokens, which can hinder accurate semantic representation of the subtasks and ultimately impact the planner's performance.\looseness=-1

To overcome these limitations, we introduce the \textbf{Pro}gressive \textbf{T}ool retrieval to \textbf{I}mprove \textbf{P}lanning (ProTIP) framework. 
Our TR strategy draws inspiration from advancements in the word embedding literature, where prior works~\cite{mikolov2013linguistic, pennington2014glove, levy2014linguistic} have shown the effectiveness of representing semantic relationships between words by embedding them in a vector space.
Extending this concept to complex queries and tools, we leverage task-specific fine-tuning to achieve our progressive TR requirements. 

ProTIP initially encodes the given query and tool descriptions to minimize the Euclidean distance between relevant tools corresponding to the first subtask and the query in a semantic space, without explicitly performing task decomposition. 
Subsequently, the ProTIP module iteratively transforms the query embedding by subtracting previously retrieved tool description embedding from the query embedding. 
The resultant embedding from this iterative subtraction aligns more closely in semantic space with a natural language query formed by eliminating previously executed subtasks from the full query, while focusing on the next most important subtask to be executed out of the remaining ones.
ProTIP is fine-tuned using contrastive loss to learn embeddings with above-mentioned characteristics, more details in section~\ref{sec:tool_retrieval}.
As a consequence, ProTIP provides flexibility by enabling incremental TR, while incorporating execution history (e.g., the previously selected tool and execution result) without the overhead of maintaining atomicity for the TD.
The contributions of this work are as follows:
\begin{itemize}
  \item We introduce ProTIP, a novel progressive TR framework, that efficiently performs TR for complex requests involving inter-subtask dependencies, while factoring in the execution history, without requiring explicit TD.
  \item We comprehensively compare various TR methods and their impact on LLM-powered planning agents using the public ToolBench dataset~\cite{toolbench:23}.
  \item To the best of our knowledge, we are the first to establish that lightweight (non-LLM) fine-tuning based tool retrieval approaches, such as ProTIP, can outperform state-of-the-art LLM-augmented approaches, such as ChatGPT-based TD for TR.
\end{itemize}

%% file: methodology.tex
\section{Data}
\label{sec:data}

\begin{figure*}[t!]
\includegraphics[width=0.5\textwidth]{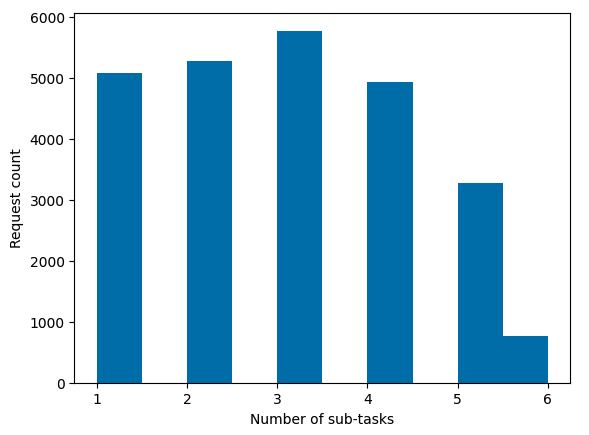} \hfill
\includegraphics[width=0.5\textwidth]{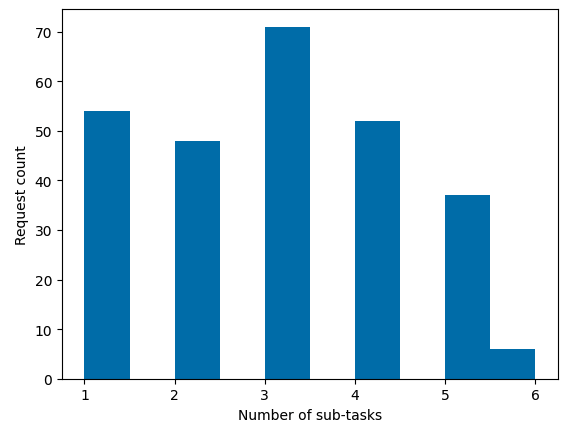}
\caption{Distribution of requests in the ToolBench training (left) and test (right) sets according to the number of subtasks involved in each request.}
\label{figure:1}
\end{figure*}

We evaluate the efficacy of ProTIP-based LLM-Planner in generating step-by-step plans using the ToolBench~\cite{toolbench:23} dataset, one of the most extensive instruction-tuning dataset for tool utilization, encompassing 16,000 tools and 49 categories. ToolBench includes 27,000 complex requests, each of which may require the use of one or more APIs/Tools to complete subtasks in a specific order. Each request is accompanied by tool and plan labels, which represent a series of more granular step-by-step instructions.

Figure~\ref{figure:1} illustrates the distribution of the number of tools required for each query, providing insights into the complexity of requests within the dataset. The analysis reveals that the maximum number of subtasks in a query is 6. This information is utilized to establish the upper bound for top-k in TR and planner experiments, as detailed in section~\ref{subsec_planner}.

\begin{figure*}[ht!]
\includegraphics[width=0.5\textwidth]{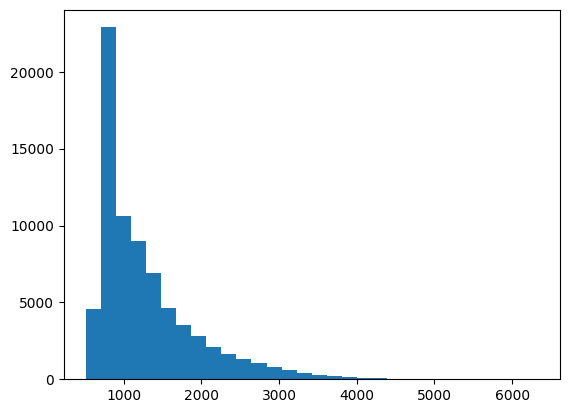} \hfill
\includegraphics[width=0.5\textwidth]{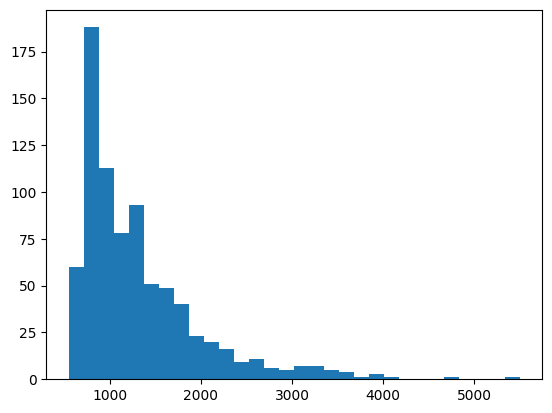}
\caption{Distributions of input token lengths for the planner in training (left) and test (right) data. The input consists of the query, top-k retrieved tools, planner-specific prompts, and execution history.}
\label{figure:2}
\end{figure*}

Figure~\ref{figure:2} shows the distribution of input token length to the planner in train and test sets. Notably, 12.25\% (97 data points) of the test set and 12.30\% (9,070 data points) of the training set exceed the context window size of 2048. This substantial proportion of lengthy inputs is expected to cause truncation, potentially hindering the model's ability to achieve optimal performance. 

\begin{figure*}[ht!]
\includegraphics[width=0.5\textwidth]{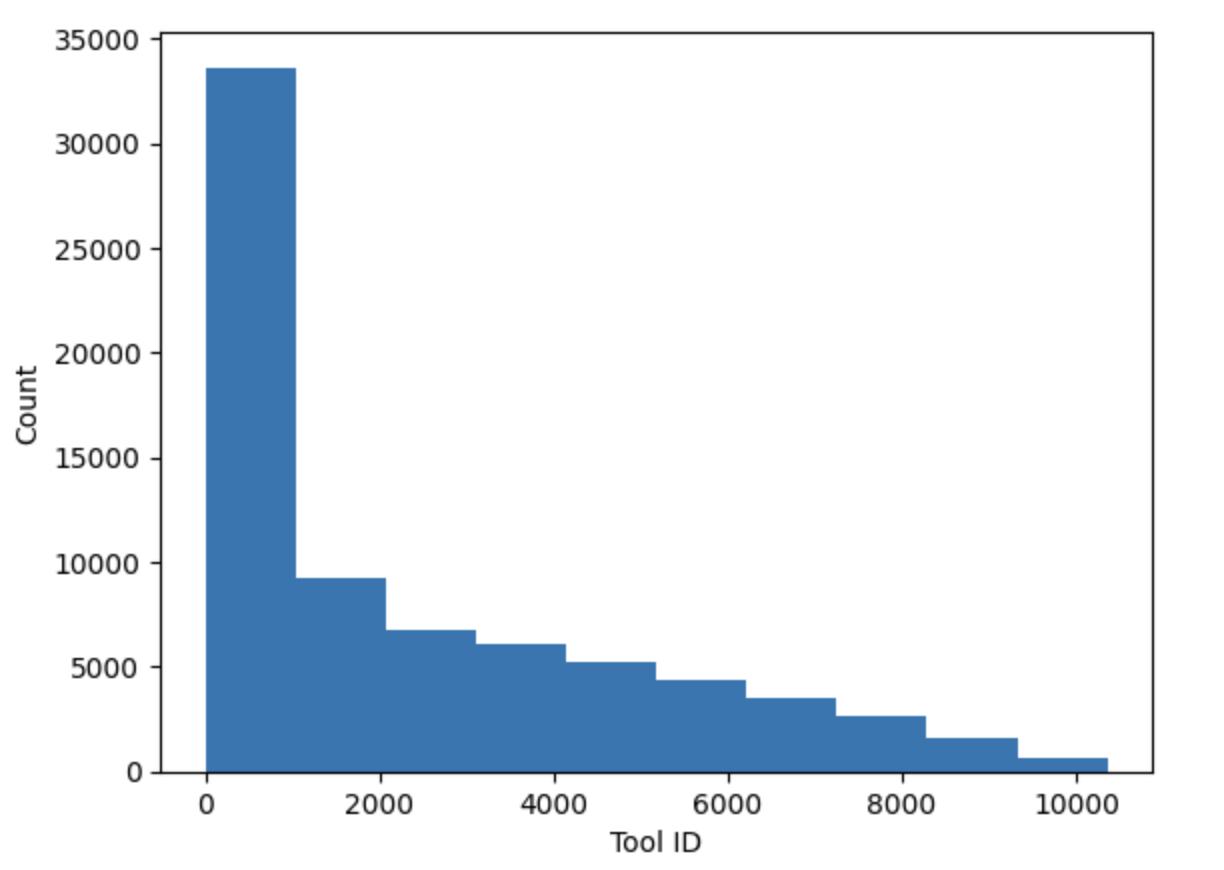} \hfill
\includegraphics[width=0.5\textwidth]{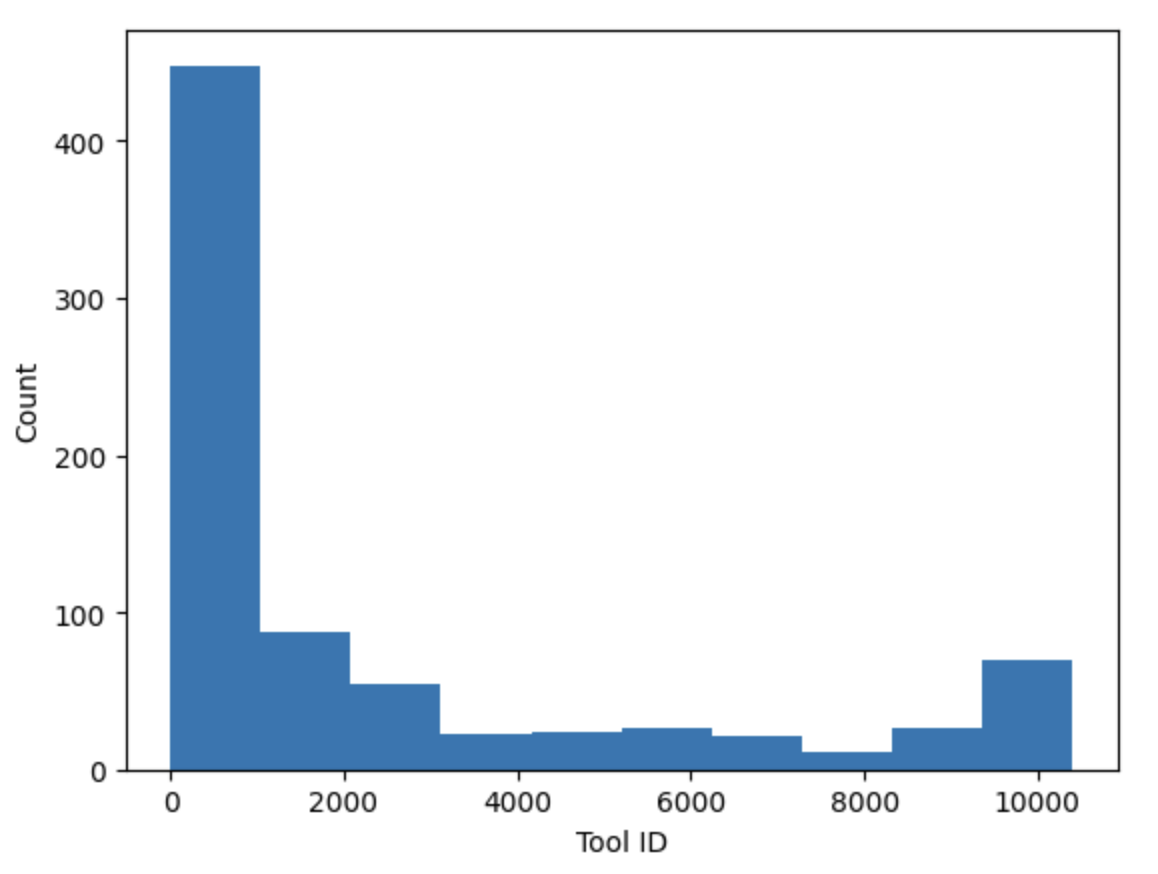}
\caption{Frequency distributions of ground truth tools in training (left) and test (right) sets. Tool names have been converted to IDs for visualization clarity.}
\label{figure:3}
\end{figure*}

Figure~\ref{figure:3} depicts the distribution of ground truth tool IDs in the dataset. Notably, a significant proportion of tool IDs fall within the range of 0 to 2000 for both the training and test sets. This imbalance in tool representation could potentially bias the model towards these more frequently appearing tools.

\begin{figure}[ht]
\centering
\includegraphics[width=\linewidth]{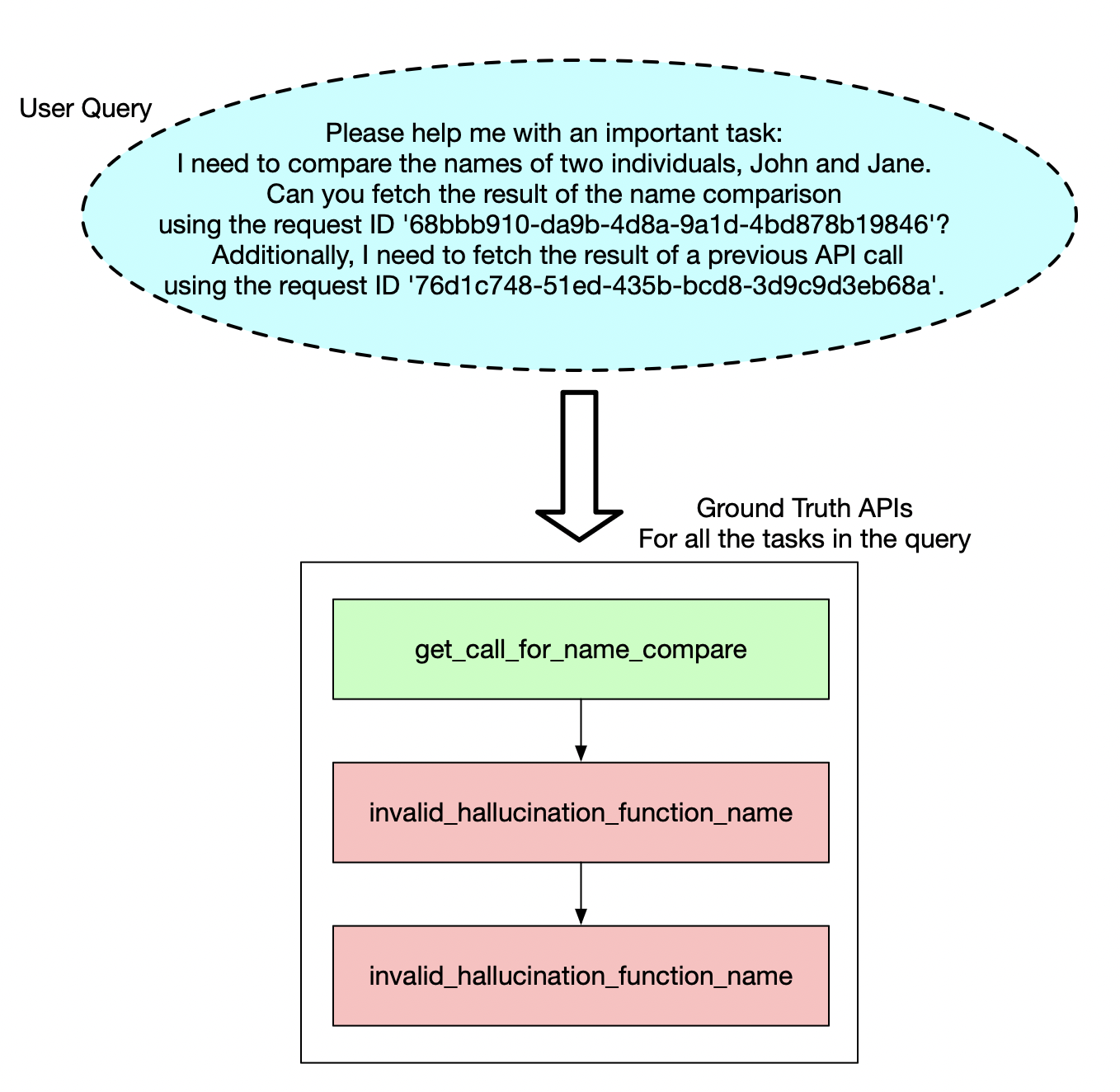}
\caption{An example of tool hallucination in the ToolBench dataset.}
\label{figure:4}
\end{figure}

The ToolBench dataset was generated using ChatGPT. As is typical with LLM-generated data without access to additional knowledge, ToolBench is susceptible to hallucinations~\cite{bang2023multitask}. An example of this can be seen in figure~\ref{figure:4}, where the synthetic dataset contains the hallucinated tool \textbf{invalid\_hallucination\_function\_name} at second and third steps. To address this issue, we removed requests with imaginary tool annotations, which are tools not included in the toolbox. Additionally, we manually reviewed and corrected remaining incorrectly extracted tools resulting from grammatical errors in the generated labels. Following the data cleaning process, the revised training and test datasets comprised of 25,489 and 274 complex requests, respectively. Using this dataset, we additionally performed TD using ChatGPT as described in section~\ref{sec:tool_retrieval}. After filtering out outliers with invalid tasks, we end up with a dataset size of 25,124 training data points and 268 test data points, which we use for all our experiments.
The average number of subtasks in our final datasets is 2.934 (Std Dev. = 1.417) for the training set and 2.955 (Std Dev. = 1.39) for the test set.

\section{Methodology}

\begin{figure*}[ht]
\includegraphics[width=\linewidth]{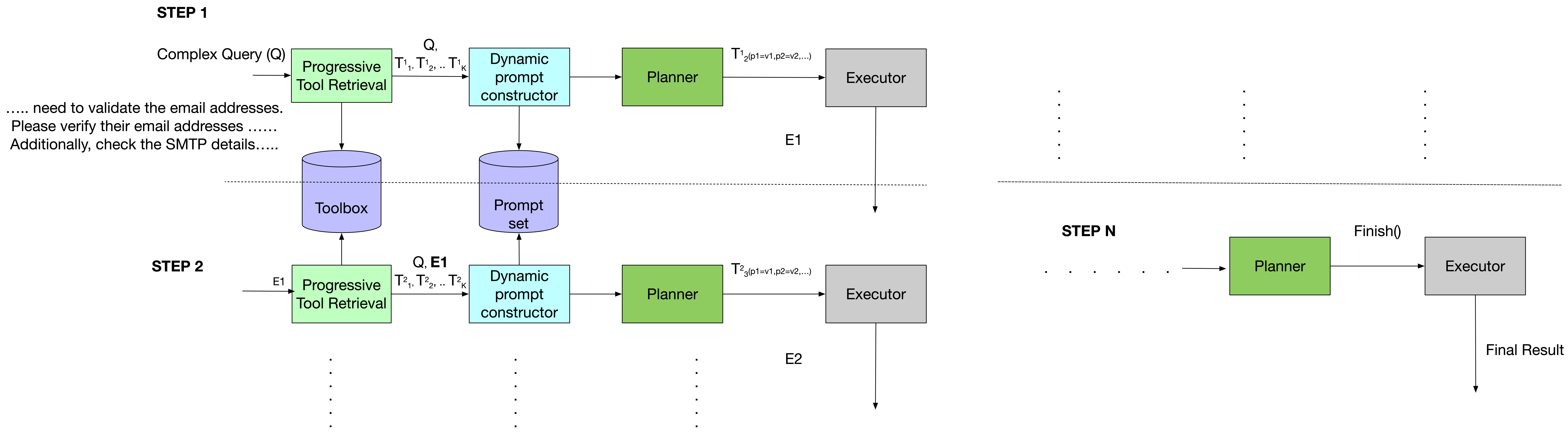}
\caption{End-to-end processing of complex queries with ProTIP-based planning.}
\label{figure:6}
\end{figure*}

To evaluate the ProTIP framework for TR and planning tasks, we use both text-based and vector-based search baselines, as well as a strong baseline of TD-based planning, on the train and test splits from the ToolBench dataset after the preprocessing step to remove bad-quality data, as described in Section~\ref{sec:data}. Figure~\ref{figure:6} illustrates the our envisioned end-to-end flow of ProTIP-based step-by-step planning.

\subsection{Tool Retrieval}
\label{sec:tool_retrieval}

Tool Retrieval (TR) aims to identify the top-k most suitable tools to empower the planner to effectively execute all subtasks within a given complex request. The toolbox $\mathrm{T} = \{(t_{1}, d_{1}), (t_{2}, d_{2}),  ..., (t_{n}, d_{n})\}$ encompasses a collection of diverse tools $t_{i}$ with predefined functionalities, captured in their tool descriptions $d_{i}$. A primary objective of TR is to extract subtasks from a complex query that align with the predefined functionalities of the tools, a concept we refer to as subtask-tool atomicity alignment and subsequently retrieve those tools.

When employing vector-based retrieval approaches, the toolbox $\mathrm{T}$ is typically represented as a vector database. An encoder $E_{w}(.)$ parameterized on $w$ produces tool description embeddings, $E_{w}(d_{i})$, which are then stored. Either the same or a query-specific encoder maps the complex query $q$ into an embedding $E_{w}(q)$. A similarity measure, such as cosine similarity, between $E_{w}(d_{i})$ and $E_{w}(q)$ is used to retrieve the top-k tools.

This study utilizes a comprehensive suite of retrieval methods, encompassing both pretrained and fine-tuned approaches, including our proposed ProTIP method, to evaluate the effectiveness of different TR strategies using the Recall@k metric.

\paragraph{\textbf{BM25}} The text-based retrieval baseline employed in this study is BM25~\cite{robertson1995okapi}. To retrieve the top-k tools, we utilize the full complex query $q$ to compute BM25 scores for each tool description $d_{i}$.

\paragraph{\textbf{Semantic Search}} For vector-based search, we utilize GTR-T5-XL~\cite{gtr-t5-xl:21} as the encoder for both query $q$ and tool descriptions $d_{i}$. The cosine similarity measure $cos\_sim(q, d_{i})$ is employed to select the top-K most relevant tools.

\paragraph{\textbf{Task Decomposition based TR}}
Task Decomposition (TD)~\cite{decomposed-prompting:2022, pet:23} breaks down a complex query $q$ into a set of subqueries $\{q_{1}, q_{2}, ..., q_{\tau}\}$, where $\tau$ denotes the number of subtasks embedded within $q$, and each $q_{i}$ represents a subquery corresponding to the $i^{th}$ subtask of $q$.

The evaluation of TD-based retrieval involves employing both BM25 and semantic search using GTR-T5-XL models. For each $q_{i}$ from TD, we perform parallel retrieval using BM25 for text-based retrieval and GTR-T5-XL for vector-based retrieval. This results in the identification of top-k tools specific to each $q_{i}$. Subsequently, we employ an interleaving strategy to determine the final top-k tools for $q$. This approach of interleaving tools with TD serves as a straightforward yet powerful baseline. We opt for tool interleaving instead of directly utilizing all subqueries simultaneously, as the top-k tools obtained using the subquery set may not align with the desired top-k tools, where each subtask effectively covers relevant tools. We use the ChatGPT~\cite{chatgpt:23} model to generate TD rewrites.

\paragraph{\textbf{ProTIP}}
We propose ProTIP, a progressive tool retrieval strategy, where top-k tools specific to each subtask are iteratively retrieved conditioned on execution history while retaining the subtask order. TD-based retrieval generates subtasks without factoring in the execution history. While TD-based retrieval can be adapted to leverage execution history, it still requires either an expensive pretrained LLM powerful enough to generate high-quality rewrites, or explicit task decomposition labels to fine-tune a lightweight model to generate rewrites. In addition, the TD labels should also ensure subtask-tool atomicity alignment is maintained.

\begin{figure}[h!]
\centering
\includegraphics[width=\linewidth]{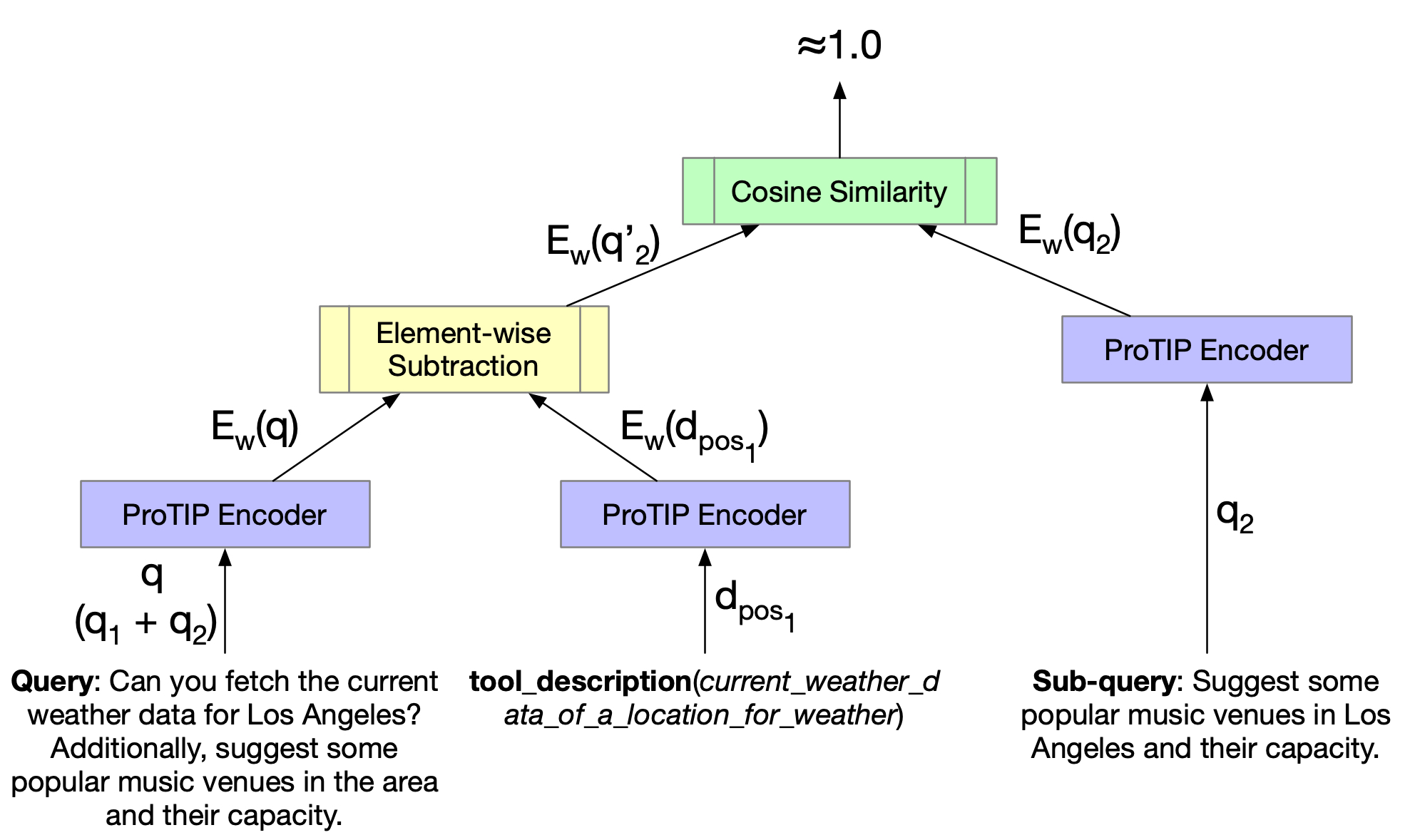}
\caption{ProTIP's implicit learning mechanism for handling complex queries. Initial retrieval selects the tool relevant to the first subtask. Subsequently, the execution history, encompassing the tool description and query, is utilized to produce a resultant embedding $E(q'_{2})$. This embedding is specifically crafted to align with $E(q_{2})$, which represents the second subtask, without the need for the subquery label $q_{2}$.}
\label{fig:ptr_imp_learning}
\end{figure}

ProTIP is a lightweight retrieval model which does not require explicit labels. Instead of relying on intermediate text representation, ProTIP directly operates in the vector space by transforming input embeddings to eliminate subtasks which are already addressed. 

Given a complex query $q$ comprised of $n$ subtasks, let $q_{1}$...$q_{n}$ denote queries that only capture each subtask from 1...$n$. We use BERT-base-uncased\footnote{\url{https://huggingface.co/bert-base-uncased}} model as the encoder for both $q$ and tool descriptions $d_{i}$, represented by $E_w(.)$. For each training batch of size $b$, we fine-tune $E_w(.)$ to always choose the ground-truth tool $t_{pos_1}$ corresponding to subtask-1 by minimizing the distance between $d_{pos_1}$ and $q$, while maximizing the distance between $q$ and a set of randomly drawn negative examples, $T_{neg} = \{t_{neg_1}, t_{neg_2}, ..., t_{neg_{b-1}}\}$, from irrelevant tools. For subsequent retrieval steps, starting with subtask-2, we iteratively subtract $E_w(t_{pos_1})$ to $E_w(t_{pos_i})$ from $E_w(q)$ to arrive at an embedding that approximates a query $q^{'}$ that only represents subtasks from $i+1$ to $n$. This operation directly results in implicit learning of task decomposition while maintaining subtask-tool atomicity without the need for explicit labels, as depicted in Figure~\ref{fig:ptr_imp_learning}.

We use contrastive loss~\cite{cl:06} to fine-tune our retrieval which is suited for metric-based learning. Given input pair with $I1$ and $I2$ inputs, contrastive loss encourages similar examples to be close, and dissimilar ones to have higher distance of at least margin $m$ from each other. We define input $I1$ for query embedding as

\begin{equation}  \label{eq:1}
  I1 = 
    \begin{cases}
      E_w(q), \text{for initial retrieval}. \\
      E_w(q) - \sum_{1 \leq i < n}(E_w(d_{i})), \text{otherwise};
    \end{cases}       
\end{equation}
where $\sum$ represents element-wise vector sum. We define tool description embedding\footnote{While we use tool descriptions, any information that helps predict the next tool could be used.} $I2$ as

\begin{equation}  \label{eq:2}
I2 = E_w(d_{j+1}),
\end{equation}
where $j \geq 0$.

The margin-based contrastive loss function is defined as

\begin{equation} \label{eq:3}
L = \frac{1}{2} l D^{2} + \frac{1}{2} (1-l) \{\max(0, m-D)\}^{2},
\end{equation}

where $l$ is a binary label which indicates whether the input pair consisting of the query $I1$ and tool description $I2$ embeddings is a positive ($l = 1$) or negative ($l = 0$) pair, m > 0 is the margin distance for dissimilar pairs and we use $0.3$. $D$ is a distance measure of choice and we use L2 norm between $I1$ and $I2$. Analogous to TD-based TR, we utilize a tool interleaving strategy to identify the final top-K tools for Recall@k evaluation.

\subsection{Progressive Planning}
\label{subsec_planner}

To retrieve tools for the progressive planning task, we employ the tool retrieval (TR) strategies proposed in Section~\ref{sec:tool_retrieval}. 
We then perform supervised fine-tuning of the OpenLLaMA-7B-v2 model~\cite{touvron2023llama, openlm2023openllama, together2023redpajama}, leveraging the HuggingFace Transformer library~\cite{huggingfacetransformers}. The model operates with a context window size of 2048 tokens. The prompt provided to the model consists of a fixed \textit{instruction}, the \textit{complex request}, and optional information such as the list of top-k \textit{API candidates} (along with their metadata) and the \textit{execution history}. This combination generates multiple distinct prompt variations.\footnote{Details in Appendix~\ref{appendix_prompt_variations}.}
In essence, our goal is to predict the next API to be executed in a multi-step plan given an input prompt containing the instruction, request, API candidates, and history. This requires unrolling the original data to form a sequence of prompts corresponding to each step in the plan.

Each interaction in the original data encompasses a natural language description of the \textit{full query}.
Additionally, each interaction comprises a total of $p$ steps labeled \textit{assistant} and $f$ steps labeled \textit{function}, along with potential inputs from the user labeled as \textit{user} (we disregard \textit{system} inputs).
To prepare training and testing data for the planner, we unroll each interaction into $p$ distinct unrolled data instances.
Within each unrolled data instance, the text generated by the \textit{assistant} for that specific step serves as the desired output, referred to as the \textit{response}, while the entire sequence of steps preceding the current step constitutes the \textit{history}.
As a general rule, we utilize the original \textit{full query} of the interaction as the \textit{request}. In the event that an input occurs amidst the steps, we simply append it to the subsequent \textit{request} segment.
Notably, the training and test data differ in terms of the tools presented as \textit{API candidates} in the input prompt.\looseness=-1

\textbf{Training:} To provide the planner with a comprehensive set of potential tools, we utilize all $p$ ground truth tools identified in the original data's \textit{assistant} steps as \textit{API candidates}. The oracle TR strategy employs the exact set of $p$ ground truth tools ($p \leq 6$) required for the entire plan in the prompt for each step, closely resembling a memorization task. In contrast, top-k TR-based planners augment the prompt with an additional ($K$ - $p$) randomly sampled tools for each step, where $K$ > $p$, alongside the $p$ ground truth tools. This approach introduces an element of uncertainty and challenges the planner to identify the most relevant tool for the current step. To ensure the correct tool is always present in the prompt, we maintain all ground truth tools from the full plan during the training of each step. This strategy guides the planner towards learning to select the most pertinent tool for the current task. Balancing between the LLM's maximum context window size of 2048 and the maximum value of $p$ (6), we set k = 10 in our experiments. To prevent the LLM from exploiting the position of the correct tool, we randomize the order of the tools presented in the prompt during training and testing.

\textbf{Testing:} In the oracle TR strategy, we use exactly $p$ ground truth tools identified from the original data's \textit{assistant} steps as \textit{API Candidates} for each step. This approach provides the Planner with a complete set of the necessary tools for each step, effectively making the task a tool selection problem. Conversely, for top-K TR-based planners, we utilize the top-10 tools retrieved by the corresponding algorithms, which may or may not include the ground truth tool. Additionally, we employ tool interleaving, where applicable.

\textbf{Evaluation:} While standard NLP metrics like Exact Match (EM)~\footnote{\url{https://github.com/huggingface/evaluate}} and ROUGELSum~\cite{lin2004rouge} are commonly used to assess the overall text quality of the generated plan, our primary focus is on evaluating the LLM's performance in selecting appropriate tools. Therefore, we employ Tool Accuracy (TA) and Tool Hallucination (TH) metrics, specifically focusing on the API component of the predicted output and disregarding the parameter details.

%% file: results.tex
\section{Results}

\subsection{Tool Retrieval}
\label{sec:tr_results}

For Recall@K, we start at K=6 given the maximum number of subtasks for a complex query in the ToolBench dataset is 6, as described in Section~\ref{sec:data}. Table~\ref{tab:tr_results} shows the recall of various retrieval approaches for different values of K.

\begin{table}[h]
\centering
\resizebox{\columnwidth}{!}{
\begin{tabular}{p{4.5cm}p{1cm}p{1cm}p{1cm}r}\\
  \toprule
Method & Recall@K \\
 \cmidrule{2-5}
 & K=6 & K=10 & K=15 & K=20 \\
 \midrule
Full query based BM25 & 31.87 & 41 & 45.72 & 48.71 \\
TD based BM25 & 41.26 & 47 & 50.70 & 54.74 \\
Full query based SS & 54.24 & 60.86 & 65.93 & 69.52 \\
TD based SS  &  57.81 &  65.57 & 69.85 & 72.8 \\
ProTIP & \textbf{80.55} & \textbf{81.36} & \textbf{82.35} & \textbf{83.48} \\
\bottomrule
\end{tabular}}
\caption{Evaluation of various tool retrieval methods on the ToolBench test set. ``TD-based" methods use task decomposition by ChatGPT and run retrieval in parallel for all subtasks, arriving at the top-K tools through interleaving. ``SS" refers to Semantic Search.}
\label{tab:tr_results}
\end{table}

\begin{figure}[h]
\centering
\includegraphics[width=\linewidth]{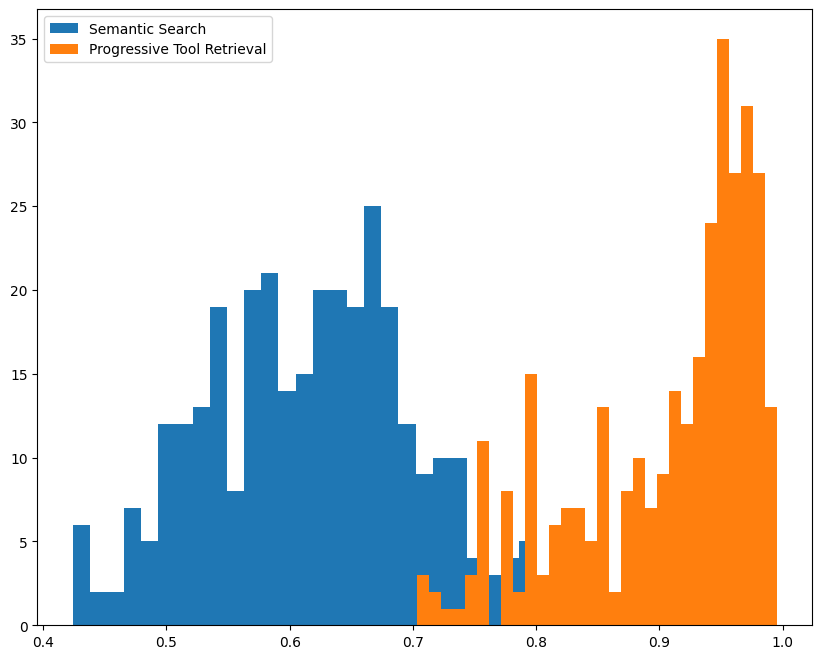}
\caption{A comparison of cosine similarity distributions between Semantic Search and Progressive Tool Retrieval. Cosine similarity was computed between ground-truth tool descriptions and complex queries from the ToolBench test data.}
\label{fig:ptr_cos_sim}
\end{figure}

Vector-based retrieval methods surpass text-based retrieval approaches, and TD-augmented retrieval employing an interleaving tools strategy demonstrates substantial gains over these baselines. ProTIP outperforms the best-performing TD-based retrieval method across all K values. As illustrated in Figure~\ref{fig:ptr_cos_sim}, ProTIP's utilization of contrastive learning enhances the cosine similarity between relevant tools and implicit subqueries. This improvement stems from iterative transformations performed directly in vector space, circumventing the requirement for intermediate text as in TD-based approaches. Consequently, ProTIP acquires implicit learning capabilities to predict the subsequent subtask and relevant tool while preserving the subtask order. The effectiveness of the ProTIP framework in handling queries characterized by complex language phenomena, such as disfluency, remains to be evaluated.

\subsection{Progressive Planning}
\label{sec:planner_results}

\begin{table*}[h]
    \centering
    \resizebox{\linewidth}{!}{
    \begin{tabular}{lcccllll}
        \hline
        \textbf{ID} & \textbf{Tool Retrieval Setting} & \textbf{Prompt} &  \textbf{EM} & \textbf{RLSum} & \textbf{TA (\%)} & \textbf{TH (\%)} \\ \hline 
        1 & BM25 with full query & [T+H] & 0.0442 & 0.3672 & 14.77 & 12.37 \\
        2 & SS with full query & [T+H] & 0.0619 & 0.4086 & \textbf{21.72} & 7.7 \\ \hline \hline
        3 & BM25 with TD query (Tool interleaving) & [T+H] & 0.053 & 0.39 & 16.29 & 8.96 \\ 
        4 & SS with TD query (Tool interleaving)  & [T+H] & 0.0833 & 0.4424 & \textbf{25.88} & 8.21 \\\hline \hline
        5 & PTR (Tool interleaving) & [T] & 0.0543 & 0.4129 & 19.82 & 2.02 \\
        6 & PTR (Tool interleaving) & [T+H] & 0.0896 & 0.4772 & \textbf{36.49} & 7.95 \\ \hline \hline
        7 & Oracle (GT + random tools) & [T] & 0.0896 & 0.5232 & 44.57 & 4.17 \\ 
        8 & Oracle (GT + random tools) & [T+H] & 0.1805 & 0.6669 & \textbf{77.53} & 17.55 \\ \hline \hline
        9 & Oracle (only GT tools) & [T] & 0.2146 & 0.579 & 46.59 & 5.3 \\ 
        10 & Oracle (only GT tools) & [T+H] & 0.3952 & 0.757 & \textbf{80.3} & 17.55 \\ \hline
    \end{tabular}
    }
    \caption{Performance of progressive plan generation using various combinations of tool retrieval algorithms and prompt generation strategies. The prompt may comprise solely the list of tools ([T]) or both history and tools ([T+H]). We present the results for scenarios where the prompt includes only the tool name as tool metadata. For a given prompt setting (i.e., [T+H]), ProTIP consistently outperforms other baseline approaches, such as BM25 and SS, both with and without task decomposition. A substantial 15.79\% absolute improvement in Recall@10 between TD-based SS and ProTIP translates to a 10.61\% absolute increase in Tool Accuracy for the Planner, accompanied by a 0.26\% reduction in Tool Hallucination.}
    \label{tab:p_planning_label}
\end{table*}

The progressive planning framework offers a multitude of configurable options, encompassing the prompt construction (with or without history, with or without candidate tools, etc.) and the type of tool metadata employed (e.g., tool name only versus tool name and description). To provide a representative assessment of the progressive planning task, we selected a subset of these configurations and evaluated the performance of various TR strategies on the preprocessed test set. The results are presented in Table~\ref{tab:p_planning_label}. Settings 1-6 utilize various TR strategies introduced in this paper, while settings 7-10 employ the oracle TR strategy. To ensure a fair comparison with full-query-based TR strategies, we adopted the interleaving strategy (detailed in Section~\ref{sec:tool_retrieval}) for generating candidate tool sets for progressive TR (PTR).

\textbf{Oracle Planner:} To assess the performance of our proposed PTR-based fine-tuned planner, we establish a benchmark using several fine-tuned Oracle planners.
These Oracle planners utilize the complete set of $p$ ground truth (GT) tools, necessary for executing the entire plan, in the prompt, mimicking the Oracle TR algorithm.
Setting 7-8 incorporates a total of 10 tools, comprising $p$ GT tools and (10 - $p$) randomly sampled tools, while setting 9-10 employs precisely the $p$ GT tools in the prompt.
Setting 9-10 can be considered a strong upper bound achievable using Oracle TR for two reasons: a) the input prompt contains all GT tools required for executing the entire plan, and b) the fine-tuned model partially memorizes the tool selection process for each step given a specific query.
We believe this represents an approximate upper bound on the performance attainable by a fine-tuned LLM-Planner employing the Oracle TR strategy, assuming the TR step achieves 100\% Recall for the tools required for each step of the full query.

\textbf{TR Planner:} Consistently outperforming other baselines like BM25 and SS, PTR demonstrates superior performance under the [T+H] prompt setting, regardless of whether TD is employed.
This superiority is further corroborated by the observed correlation between Recall@K of the TR algorithm (BM25 < SS < PTR) and Tool Accuracy (TA) of the Planner.
Additionally, the better performance of BM25 and SS-based TR for task-decomposed queries is reflected in the corresponding Planner performance.
This aligns with the Planner's design, which mandates tool selection from the TR algorithm's retrieved set.
Interestingly, the Tool Hallucination (TH) percentage, which represents the proportion of times the Planner creates non-existent tools, reveals a consequence of failing to adhere to this design principle.
PTR without history exhibits the lowest TH percentage, despite its relatively low TA.
Upon incorporating history, both PTR (setting 6) and Oracle (settings 8 and 10) experience an increase in TA and TH, potentially due to truncation issues (discussed in Section~\ref{sec:limitations}).
Notably, higher TA in PTR leads to marginal improvements in Exact Match (EM) and ROUGELSum (RLSum), metrics that evaluate the entire predicted output, including tool parameters.
The relatively modest gains in these metrics suggest that further performance enhancements can be achieved by focusing on tool parameter optimization.
The performance gap between Oracle planners (settings 6 to 10) and the PTR-based planner highlights the potential for further Planner performance improvements.

\textbf{Importance of history for Planning:} The inclusion of the history of previous steps demonstrates a significant performance boost in planning across all settings. This improvement is particularly pronounced for both Oracle-based planning (approx. 70+\% improvement between settings 9 and 10) and PTR-based planning (approx. 80+\% improvement between settings 5 and 6) in TA. Intuitively, incorporating history is crucial as it can aid in selecting the most appropriate tool, especially during branching scenarios that may arise during the execution of the previous tool (e.g., if the previous tool executed successfully, select Tool-A; otherwise, select Tool-B). However, incorporating history into the prompt raises concerns about truncation due to the increased token count (more details in Section~\ref{sec:limitations}).

\section{Limitations and Discussion}
\label{sec:limitations}

Due to the computational demands of hyperparameter tuning, we were unable to optimize the settings for all configurations. Each configuration requires 8 A100 GPUs on AWS, resulting in significant time and resource consumption. Consequently, we focused our hyperparameter tuning efforts on the ProTIP (settings 5 and 6) and Oracle (settings 9 and 10). The detailed hyperparameter values for all settings in Table~\ref{tab:p_planning_label} are provided in Appendix~\ref{appendix_hyperparameters}. To ensure a fair comparison with the full query-based TR strategies, we employed the interleaving strategy (described in Section~\ref{sec:tool_retrieval}) for generating candidate tool sets for PTR. We recognize that this approach is not ideal for evaluating the planner's performance under PTR, as the optimal approach would involve retrieving tools step-by-step and performing planning at each step. However, this would require a mechanism to execute the predicted API call at each step and incorporate the resulting output into the planning process for the next step. While we are currently investigating potential solutions for this challenge, planner design is not the primary focus of this work. Therefore, we defer the development and evaluation of end-to-end step-by-step planning experiments, including performance tuning, to future research.

The experiment results reveal a substantial performance gap between the Oracle planner and the TR-based planners. This disparity can be attributed to two key factors. First, the Oracle planner (settings 9 and 10) utilizes the exact set of $p$ ground truth tools specified in the prompt for each progressive planning step ($p \leq 6$), whereas the TR planners operate on a larger candidate set of K=10 tools. This restricted tool selection for the Oracle planner (settings 9 and 10) likely contributes to its improved performance. This observation is further supported by the higher TA achieved in setting 10 (using exactly $p$ ground truth tools) compared to setting 8 (using K tools, with $p$ ground truth tools and (10 - $p$) randomly sampled tools).

The tool distribution discrepancy between the train and test sets, particularly for tool IDs greater than 8000, as evident in Figure~\ref{figure:3}, may partially explain the inferior performance of all retrieval-based planners. This disparity in tool distribution could hinder the effectiveness of the TR strategies, potentially leading to suboptimal planning decisions, unless tool metadata is further enriched and included in the prompt during training to support for better generalization. Additionally, we observed a poor Accuracy for the special \textit{Finish} tool, which resulted in overall low performance in all the TR settings.

The training strategy of employing the $p$ ground truth tools alongside ($K$ - $p$) randomly sampled tools in the prompt may contribute to the lower performance of the TR planner models. The presence of the ground truth tools alongside semantically dissimilar randomly sampled tools during training likely facilitates the models' ability to identify the correct tool. However, during testing, the prompt comprises top-K tools retrieved by the TR algorithms, which may exhibit semantic similarity to the ground truth tool. This semantic similarity poses a challenge for the models during inference, leading to the observed low TA values for all TR-based planner models. Utilizing the top-K tools retrieved by the TR algorithms during training could exacerbate this issue, as there is a risk of the prompt not containing the correct tool for the corresponding step. This would further complicate the learning process for the LLM and increase the likelihood of hallucinations.

To address this limitation, in future, an alternative training data creation strategy could be employed. Instead of using randomly sampled tools, the training data could incorporate tools retrieved by the TR algorithm on the training set. Additionally, to ensure that the training process effectively addresses all steps, the ground truth tool for the current step could be injected into the prompt if it is not already present. By adopting this modified training approach, we aim to enhance the performance of the TR planner models and improve their generalization capabilities. The Instructions part of the prompt are generic and can be further modified (i.e., made more precise for each scenario) to be more specific to various prompt settings. Also, we did not conduct an exhaustive study to test the robustness of the planner output to different types of input prompt variations (e.g.: paraphrased query as inputs, semantically similar tools, unseen tools in the test set etc.), which we leave as future work.

Our experiments highlight the significance of the \textit{history} in achieving higher TA for both Oracle (setting 9 vs. 10) and PTR (setting 5 vs. 6) based planning strategies. However, we believe that TA can be further improved while reducing TH, particularly for TR planners with K=10 tools, as the \textit{history} contributes to the long context issue. We observe that for the scenarios where the input size becomes close to the maximum context window size, the LLM could generate empty plans, which contributes to 3\% to 5\% of the errors in our experiments, thereby negatively impacting the TA. Note that the original data contains intermediate steps with verbose outputs that provide minimal contextual information (e.g., weather API outputs with latitude, longitude, last updated time, etc.), all of which may not be essential for determining the next API. Preserving these verbose outputs in the \textit{history} exacerbates the truncation problem, thereby negatively impacting the learning and plan generation capability of the model. This issue can be further aggravated by incorporating more tool metadata (tool description, parameters, API signature, etc.) into the prompt, which will increase the input length of the prompt even further. However for better generalization to unseen tools, ideally we want to incorporate such additional metadata into the prompt, which requires further investigation.\looseness=-1

Increasing the context window size of LLMs (e.g., to 4096 or higher) or employing techniques that allow for larger input text (e.g., as proposed in ~\cite{beltagy2020longformer}) can largely alleviate the truncation problem. However, even with large context window, studies by ~\cite{liu2023lost} indicate that LLMs tend to focus on information at the beginning or end of the prompt, even with a large context window. Therefore, alongside exploring LLMs with larger context windows, there is a need to develop methods for effectively compressing and presenting relevant contextual information, particularly the \textit{history}, to the LLM~\cite{ge2023context} to enhance performance. 

In the current work, we focused heavily on the tool accuracy across the tool retrieval and planning steps~\cite{patil2023gorilla}. Tool parameter accuracy is another important aspect of the planner output~\cite{shen2023taskbench}, which requires further investigations to improve the performance. We did not conduct any safety study or red-teaming to evaluate the bias or risks emanating from the outputs generated by the fine-tuned LLM. We want to refer to a contemporary work by~\cite{valmeekam2023planning} which has pointed out that the ability of LLM's to generate ``executable plans'' in a completely autonomous way is very limited. In our work, while planning is not the primary focus, we observed that plan generation using supervised fine-tuning of a LLM is not an easy task, specifically with a relatively small LLM (e.g.: LLM with 7B parameters). We believe further research is required to enhance our understanding of the true capabilities of LLM's for the planning task.

%% file: related_work.tex
\section{Related Work}

\textbf{Tool Retrieval using Neural embedding:} Vector databases enable storing tool name and description embeddings generated by an encoding model~\cite{cer2018universal}. These embeddings are then leveraged for semantic similarity computation, utilizing measures like cosine similarity, with queries or sub-queries. Building on the established approach of utilizing neural networks to generate task-specific semantic word/sentence embeddings for information retrieval and NLP tasks~\cite{zhang2016neural}, this work proposes a tool embedding generation strategy specifically designed to facilitate step-by-step planning.

Word embeddings~\cite{mikolov-etal-2013-linguistic, pennington2014glove, levy2014linguistic, jiao2021brief}, learned vectors representing various linguistic and semantic aspects of words, have revolutionized Natural Language Processing by enabling efficient solutions to diverse tasks like analogy queries~\cite{levy2014linguistic, allen2019analogies}. Building upon this success, research has extended to generating sentence, paragraph, and document-level embeddings~\cite{le2014distributed,wieting2015towards,li2015hierarchical} for various applications. Similarly, the Knowledge Graph domain utilizes node embedding to encode entity relationships, trained with custom objective functions to capture latent relationships in the vector space for subsequent exploitation~\cite{wang8047276}. We leverage this paradigm, employing progressive tool retrieval with fine-tuned embeddings optimized for the step-by-step planning task.

\textbf{LLM as Planner:} LLMs have emerged as potent few-shot learners~\cite{brown2020language:20, rae2022scaling}, exhibiting remarkable prowess across diverse language tasks. However, planning remains a challenging research frontier despite their impressive performance in these domains. Planning involves decomposing a high-level natural language (NL) task into a sequence of executable steps realizable by an agent. This process demands both language comprehension and an understanding of a predefined set of actions, tools, APIs, and other grounding elements. In the realm of embodied agents, LLMs have been harnessed to decompose NL instructions into simpler, more manageable units~\cite{huang2022language, ahn2022i, singh2022progprompt, decomposed-prompting:2022, pet:23, shen2023taskbench}. Notably, using LLMs to generate tool/API calls as part of the planning process can be akin to multi-step program synthesis~\cite{li2023starcoder, nijkamp2022codegen,nijkamp2023codegen2}. More recent works have tried to further improve LLM performance by adding the capability to reason/criticize the LLM outputs~\cite{kim2023language, yao2022react}.

While contemporary research has emphasized leveraging tools to enhance LLM capabilities, most existing retrieval systems rely on vector databases, similar to the renowned Retrieval Augmented Generation (RAG) technique~\cite{lewis2021retrievalaugmented}, to store and retrieve non-parametric knowledge absent in the LLM. Recent work has explored individual tools like web search engines~\cite{nakano2022webgpt}, calculators~\cite{andor2019giving}, and generic toolsets~\cite{schick2023toolformer} for planning, while others have integrated LLMs with expansive API collections to address more open-ended tasks~\cite{patil2023gorilla, shen2023hugginggpt, liang2023taskmatrixai, toolbench:23}. Fine-tuning with tool-specific data is often employed to improve task performance. However, as the number of tool grows, retrieval-based systems emerge as an efficient means for selecting the relevant tools for a given request~\cite{patil2023gorilla, toolbench:23}. Building upon this paradigm, our approach proposes the novel concept of incrementally generating tool candidates specific to the current sub-step within a multi-step planning task, ultimately enhancing the LLM's overall planning performance.

%% file: conclusion.tex
\section{Conclusion}
We introduce ProTIP, a novel lightweight tool retrieval framework, which surpasses LLM-based planning agents equipped with state-of-the-art task decomposition retrieval powered by ChatGPT. 
ProTIP's iterative vector transformations, enabled by contrastive learning, facilitate implicit learning of sequential subtask prediction, eliminating the need for explicit subtask labels. Additionally, ProTIP effectively handles "subtask-tool atomicity alignment."
On the ToolBench dataset, ProTIP framework surpasses the ChatGPT task decomposition based approach by 24\% on Recall@K=10 for Tool Retrieval and by 41\% on Tool Accuracy for plan generation.

\section{Acknowledgements}
We would like to thank Stephen Pulman, Russ Webb and Arturo Argueta for their valuable feedback. Also we thank Jiarui Lu, Yuan Zhang, Xuan Wang, Hans Han, and Jian Zhang for providing infrastructure support to fine-tune LLMs.

%% file: appendix.tex
\appendix

\section{Prompt Variations}
\label{appendix_prompt_variations}

\begin{figure}[htbp]
\centering
\includegraphics[width=\textwidth]{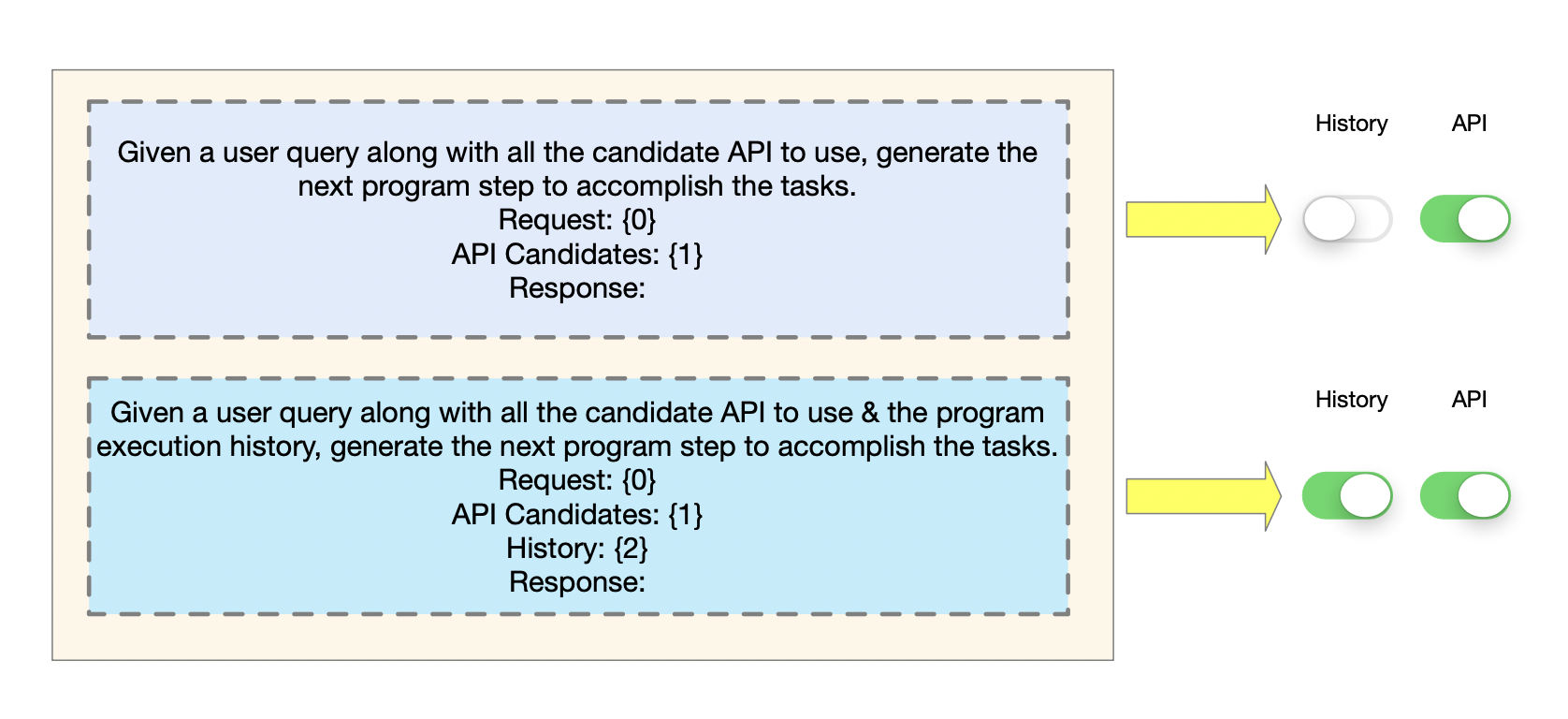}
\caption{Prompt variants for plan generation with and without execution history.}
\label{fig:prompt_variations}
\end{figure}

\section{Hyperparemeters}
\label{appendix_hyperparameters}

Due to time and resource constraints, we were unable to perform hyperparameter tuning for all experimental setups. The following table lists the hyperparameters used for each of the settings.

\begin{table}[!htb]
\caption{Hyperparameters for training}
\label{tab:planning_hyperparameters}
\centering
\resizebox{\linewidth}{!}{
\begin{tabular}{p{1cm}p{1cm}p{1cm}p{1cm}p{1.5cm}}\\
\hline
Setting ID & Epochs & LR & Batch & Warmup Steps\\
\hline
1 & 3 & 3e-5 & 64 & 20 \\
2 & 3 & 3e-5 & 64 & 20 \\
3 & 3 & 3e-5 & 64 & 20 \\
4 & 3 & 3e-5 & 64 & 20 \\
5 & 3 & 3e-5 & 64 & 20 \\
6 & 3 & 5e-5 & 128 & 20 \\
7 & 3 & 5e-5 & 128 & 20 \\
8 & 3 & 5e-5 & 128 & 20 \\
9 & 3 & 2e-5 & 64 & 20 \\
10 & 3 & 2e-5 & 64 & 20 \\
\bottomrule
\end{tabular}}
\end{table}

This model utilizes a maximum context window size of 2048 and employs a weight decay of 0 across all experiments.

%% file: naacl2021.bbl
\begin{thebibliography}{63}
\expandafter\ifx\csname natexlab\endcsname\relax\def\natexlab#1{#1}\fi

\bibitem[{Ahn et~al.(2022)Ahn, Brohan, Brown, Chebotar, Cortes, David, Finn, Fu, Gopalakrishnan, Hausman, Herzog, Ho, Hsu, Ibarz, Ichter, Irpan, Jang, Ruano, Jeffrey, Jesmonth, Joshi, Julian, Kalashnikov, Kuang, Lee, Levine, Lu, Luu, Parada, Pastor, Quiambao, Rao, Rettinghouse, Reyes, Sermanet, Sievers, Tan, Toshev, Vanhoucke, Xia, Xiao, Xu, Xu, Yan, and Zeng}]{ahn2022i}
Michael Ahn, Anthony Brohan, Noah Brown, Yevgen Chebotar, Omar Cortes, Byron David, Chelsea Finn, Chuyuan Fu, Keerthana Gopalakrishnan, Karol Hausman, Alex Herzog, Daniel Ho, Jasmine Hsu, Julian Ibarz, Brian Ichter, Alex Irpan, Eric Jang, Rosario~Jauregui Ruano, Kyle Jeffrey, Sally Jesmonth, Nikhil~J Joshi, Ryan Julian, Dmitry Kalashnikov, Yuheng Kuang, Kuang-Huei Lee, Sergey Levine, Yao Lu, Linda Luu, Carolina Parada, Peter Pastor, Jornell Quiambao, Kanishka Rao, Jarek Rettinghouse, Diego Reyes, Pierre Sermanet, Nicolas Sievers, Clayton Tan, Alexander Toshev, Vincent Vanhoucke, Fei Xia, Ted Xiao, Peng Xu, Sichun Xu, Mengyuan Yan, and Andy Zeng. 2022.
\newblock \href {http://arxiv.org/abs/2204.01691} {Do as i can, not as i say: Grounding language in robotic affordances}.

\bibitem[{Allen and Hospedales(2019)}]{allen2019analogies}
Carl Allen and Timothy Hospedales. 2019.
\newblock \href {http://arxiv.org/abs/1901.09813} {Analogies explained: Towards understanding word embeddings}.

\bibitem[{Andor et~al.(2019)Andor, He, Lee, and Pitler}]{andor2019giving}
Daniel Andor, Luheng He, Kenton Lee, and Emily Pitler. 2019.
\newblock \href {http://arxiv.org/abs/1909.00109} {Giving bert a calculator: Finding operations and arguments with reading comprehension}.

\bibitem[{Bang et~al.(2023)Bang, Cahyawijaya, Lee, Dai, Su, Wilie, Lovenia, Ji, Yu, Chung et~al.}]{bang2023multitask}
Yejin Bang, Samuel Cahyawijaya, Nayeon Lee, Wenliang Dai, Dan Su, Bryan Wilie, Holy Lovenia, Ziwei Ji, Tiezheng Yu, Willy Chung, et~al. 2023.
\newblock A multitask, multilingual, multimodal evaluation of chatgpt on reasoning, hallucination, and interactivity.
\newblock \emph{arXiv preprint arXiv:2302.04023}.

\bibitem[{Beltagy et~al.(2020)Beltagy, Peters, and Cohan}]{beltagy2020longformer}
Iz~Beltagy, Matthew~E Peters, and Arman Cohan. 2020.
\newblock Longformer: The long-document transformer.
\newblock \emph{arXiv preprint arXiv:2004.05150}.

\bibitem[{Brown et~al.(2020)Brown, Mann, Ryder, Subbiah, Kaplan, Dhariwal, Neelakantan, Shyam, Sastry, Askell et~al.}]{brown2020language:20}
Tom Brown, Benjamin Mann, Nick Ryder, Melanie Subbiah, Jared~D Kaplan, Prafulla Dhariwal, Arvind Neelakantan, Pranav Shyam, Girish Sastry, Amanda Askell, et~al. 2020.
\newblock Language models are few-shot learners.
\newblock \emph{Advances in neural information processing systems}, 33:1877--1901.

\bibitem[{Bubeck et~al.(2023)Bubeck, Chandrasekaran, Eldan, Gehrke, Horvitz, Kamar, Lee, Lee, Li, Lundberg et~al.}]{bubeck2023sparks}
S{\'e}bastien Bubeck, Varun Chandrasekaran, Ronen Eldan, Johannes Gehrke, Eric Horvitz, Ece Kamar, Peter Lee, Yin~Tat Lee, Yuanzhi Li, Scott Lundberg, et~al. 2023.
\newblock Sparks of artificial general intelligence: Early experiments with gpt-4.
\newblock \emph{arXiv preprint arXiv:2303.12712}.

\bibitem[{Cer et~al.(2018)Cer, Yang, yi~Kong, Hua, Limtiaco, John, Constant, Guajardo-Cespedes, Yuan, Tar, Sung, Strope, and Kurzweil}]{cer2018universal}
Daniel Cer, Yinfei Yang, Sheng yi~Kong, Nan Hua, Nicole Limtiaco, Rhomni~St. John, Noah Constant, Mario Guajardo-Cespedes, Steve Yuan, Chris Tar, Yun-Hsuan Sung, Brian Strope, and Ray Kurzweil. 2018.
\newblock \href {http://arxiv.org/abs/1803.11175} {Universal sentence encoder}.

\bibitem[{Chiang et~al.(2023)Chiang, Li, Lin, Sheng, Wu, Zhang, Zheng, Zhuang, Zhuang, Gonzalez, Stoica, and Xing}]{vicuna2023}
Wei-Lin Chiang, Zhuohan Li, Zi~Lin, Ying Sheng, Zhanghao Wu, Hao Zhang, Lianmin Zheng, Siyuan Zhuang, Yonghao Zhuang, Joseph~E. Gonzalez, Ion Stoica, and Eric~P. Xing. 2023.
\newblock \href {https://lmsys.org/blog/2023-03-30-vicuna/} {Vicuna: An open-source chatbot impressing gpt-4 with 90\%* chatgpt quality}.

\bibitem[{Chowdhery et~al.(2022)Chowdhery, Narang, Devlin, Bosma, Mishra, Roberts, Barham, Chung, Sutton, Gehrmann et~al.}]{chowdhery2022palm}
Aakanksha Chowdhery, Sharan Narang, Jacob Devlin, Maarten Bosma, Gaurav Mishra, Adam Roberts, Paul Barham, Hyung~Won Chung, Charles Sutton, Sebastian Gehrmann, et~al. 2022.
\newblock Palm: Scaling language modeling with pathways.
\newblock \emph{arXiv preprint arXiv:2204.02311}.

\bibitem[{Computer(2023)}]{together2023redpajama}
Together Computer. 2023.
\newblock \href {https://github.com/togethercomputer/RedPajama-Data} {Redpajama-data: An open source recipe to reproduce llama training dataset}.

\bibitem[{Ge et~al.(2023)Ge, Hu, Wang, Chen, and Wei}]{ge2023context}
Tao Ge, Jing Hu, Xun Wang, Si-Qing Chen, and Furu Wei. 2023.
\newblock In-context autoencoder for context compression in a large language model.
\newblock \emph{arXiv preprint arXiv:2307.06945}.

\bibitem[{Geng and Liu(2023)}]{openlm2023openllama}
Xinyang Geng and Hao Liu. 2023.
\newblock \href {https://github.com/openlm-research/open_llama} {Openllama: An open reproduction of llama}.

\bibitem[{Hadsell et~al.(2006)Hadsell, Chopra, and LeCun}]{cl:06}
R.~Hadsell, S.~Chopra, and Y.~LeCun. 2006.
\newblock \href {https://ieeexplore.ieee.org/document/1640964} {Dimensionality reduction by learning an invariant mapping}.
\newblock In \emph{Proceedings of 2006 IEEE Computer Society Conference on Computer Vision and Pattern Recognition (CVPR'06).}

\bibitem[{Huang et~al.(2022)Huang, Abbeel, Pathak, and Mordatch}]{huang2022language}
Wenlong Huang, Pieter Abbeel, Deepak Pathak, and Igor Mordatch. 2022.
\newblock \href {http://arxiv.org/abs/2201.07207} {Language models as zero-shot planners: Extracting actionable knowledge for embodied agents}.

\bibitem[{Jiao and Zhang(2021)}]{jiao2021brief}
Qilu Jiao and Shunyao Zhang. 2021.
\newblock A brief survey of word embedding and its recent development.
\newblock In \emph{2021 IEEE 5th Advanced Information Technology, Electronic and Automation Control Conference (IAEAC)}, volume~5, pages 1697--1701. IEEE.

\bibitem[{Khot et~al.(2022)Khot, Trivedi, Finlayson, Fu, Richardson, Clark, and Sabharwal}]{decomposed-prompting:2022}
Tushar Khot, Harsh Trivedi, Matthew Finlayson, Yao Fu, Kyle Richardson, Peter Clark, and Ashish Sabharwal. 2022.
\newblock Decomposed prompting: A modular approach for solving complex tasks.
\newblock In \emph{Proceedings of the Eleventh International Conference on Learning Representations}.

\bibitem[{Kim et~al.(2023)Kim, Baldi, and McAleer}]{kim2023language}
Geunwoo Kim, Pierre Baldi, and Stephen McAleer. 2023.
\newblock Language models can solve computer tasks.
\newblock \emph{arXiv preprint arXiv:2303.17491}.

\bibitem[{Kojima et~al.(2022)Kojima, Gu, Reid, Matsuo, and Iwasawa}]{kojima2022large}
Takeshi Kojima, Shixiang~Shane Gu, Machel Reid, Yutaka Matsuo, and Yusuke Iwasawa. 2022.
\newblock Large language models are zero-shot reasoners.
\newblock \emph{Advances in neural information processing systems}, 35:22199--22213.

\bibitem[{Le and Mikolov(2014)}]{le2014distributed}
Quoc Le and Tomas Mikolov. 2014.
\newblock Distributed representations of sentences and documents.
\newblock In \emph{International conference on machine learning}, pages 1188--1196. PMLR.

\bibitem[{Levy and Goldberg(2014)}]{levy2014linguistic}
Omer Levy and Yoav Goldberg. 2014.
\newblock Linguistic regularities in sparse and explicit word representations.
\newblock In \emph{Proceedings of the eighteenth conference on computational natural language learning}, pages 171--180.

\bibitem[{Lewis et~al.(2021)Lewis, Perez, Piktus, Petroni, Karpukhin, Goyal, Küttler, Lewis, tau Yih, Rocktäschel, Riedel, and Kiela}]{lewis2021retrievalaugmented}
Patrick Lewis, Ethan Perez, Aleksandra Piktus, Fabio Petroni, Vladimir Karpukhin, Naman Goyal, Heinrich Küttler, Mike Lewis, Wen tau Yih, Tim Rocktäschel, Sebastian Riedel, and Douwe Kiela. 2021.
\newblock \href {http://arxiv.org/abs/2005.11401} {Retrieval-augmented generation for knowledge-intensive nlp tasks}.

\bibitem[{Li et~al.(2015)Li, Luong, and Jurafsky}]{li2015hierarchical}
Jiwei Li, Minh-Thang Luong, and Dan Jurafsky. 2015.
\newblock A hierarchical neural autoencoder for paragraphs and documents.
\newblock \emph{arXiv preprint arXiv:1506.01057}.

\bibitem[{Li et~al.(2023)Li, Allal, Zi, Muennighoff, Kocetkov, Mou, Marone, Akiki, Li, Chim et~al.}]{li2023starcoder}
Raymond Li, Loubna~Ben Allal, Yangtian Zi, Niklas Muennighoff, Denis Kocetkov, Chenghao Mou, Marc Marone, Christopher Akiki, Jia Li, Jenny Chim, et~al. 2023.
\newblock Starcoder: may the source be with you!
\newblock \emph{arXiv preprint arXiv:2305.06161}.

\bibitem[{Liang et~al.(2023)Liang, Wu, Song, Wu, Xia, Liu, Ou, Lu, Ji, Mao, Wang, Shou, Gong, and Duan}]{liang2023taskmatrixai}
Yaobo Liang, Chenfei Wu, Ting Song, Wenshan Wu, Yan Xia, Yu~Liu, Yang Ou, Shuai Lu, Lei Ji, Shaoguang Mao, Yun Wang, Linjun Shou, Ming Gong, and Nan Duan. 2023.
\newblock \href {http://arxiv.org/abs/2303.16434} {Taskmatrix.ai: Completing tasks by connecting foundation models with millions of apis}.

\bibitem[{Lin(2004)}]{lin2004rouge}
Chin-Yew Lin. 2004.
\newblock \href {https://www.aclweb.org/anthology/W04-1013} {{ROUGE}: A package for automatic evaluation of summaries}.
\newblock In \emph{Text Summarization Branches Out}, pages 74--81, Barcelona, Spain. Association for Computational Linguistics.

\bibitem[{Liu et~al.(2023)Liu, Lin, Hewitt, Paranjape, Bevilacqua, Petroni, and Liang}]{liu2023lost}
Nelson~F Liu, Kevin Lin, John Hewitt, Ashwin Paranjape, Michele Bevilacqua, Fabio Petroni, and Percy Liang. 2023.
\newblock Lost in the middle: How language models use long contexts.
\newblock \emph{arXiv preprint arXiv:2307.03172}.

\bibitem[{Mikolov et~al.(2013{\natexlab{a}})Mikolov, Yih, and Zweig}]{mikolov2013linguistic}
Tom{\'a}{\v{s}} Mikolov, Wen-tau Yih, and Geoffrey Zweig. 2013{\natexlab{a}}.
\newblock Linguistic regularities in continuous space word representations.
\newblock In \emph{Proceedings of the 2013 conference of the north american chapter of the association for computational linguistics: Human language technologies}, pages 746--751.

\bibitem[{Mikolov et~al.(2013{\natexlab{b}})Mikolov, Yih, and Zweig}]{mikolov-etal-2013-linguistic}
Tomas Mikolov, Wen-tau Yih, and Geoffrey Zweig. 2013{\natexlab{b}}.
\newblock \href {https://aclanthology.org/N13-1090} {Linguistic regularities in continuous space word representations}.
\newblock In \emph{Proceedings of the 2013 Conference of the North {A}merican Chapter of the Association for Computational Linguistics: Human Language Technologies}, pages 746--751, Atlanta, Georgia. Association for Computational Linguistics.

\bibitem[{Min et~al.(2022)Min, Lyu, Holtzman, Artetxe, Lewis, Hajishirzi, and Zettlemoyer}]{minetal2022rethinking}
Sewon Min, Xinxi Lyu, Ari Holtzman, Mikel Artetxe, Mike Lewis, Hannaneh Hajishirzi, and Luke Zettlemoyer. 2022.
\newblock Rethinking the role of demonstrations: What makes in-context learning work?
\newblock In \emph{Proceedings of the 2022 Conference on Empirical Methods in Natural Language Processing}.

\bibitem[{Nakano et~al.(2022)Nakano, Hilton, Balaji, Wu, Ouyang, Kim, Hesse, Jain, Kosaraju, Saunders, Jiang, Cobbe, Eloundou, Krueger, Button, Knight, Chess, and Schulman}]{nakano2022webgpt}
Reiichiro Nakano, Jacob Hilton, Suchir Balaji, Jeff Wu, Long Ouyang, Christina Kim, Christopher Hesse, Shantanu Jain, Vineet Kosaraju, William Saunders, Xu~Jiang, Karl Cobbe, Tyna Eloundou, Gretchen Krueger, Kevin Button, Matthew Knight, Benjamin Chess, and John Schulman. 2022.
\newblock \href {http://arxiv.org/abs/2112.09332} {Webgpt: Browser-assisted question-answering with human feedback}.

\bibitem[{Ni et~al.(2021)Ni, Qu, Lu, Dai, Ábrego, Ma, Zhao, Luan, Hall, Chang, and Yang}]{gtr-t5-xl:21}
Jianmo Ni, Chen Qu, Jing Lu, Zhuyun Dai, Gustavo~Hernández Ábrego, Ji~Ma, Vincent~Y. Zhao, Yi~Luan, Keith~B. Hall, Ming-Wei Chang, and Yinfei Yang. 2021.
\newblock \href {https://arxiv.org/abs/2112.07899} {Large dual encoders are generalizable retrievers}.

\bibitem[{Nijkamp et~al.(2023)Nijkamp, Hayashi, Xiong, Savarese, and Zhou}]{nijkamp2023codegen2}
Erik Nijkamp, Hiroaki Hayashi, Caiming Xiong, Silvio Savarese, and Yingbo Zhou. 2023.
\newblock Codegen2: Lessons for training llms on programming and natural languages.
\newblock \emph{arXiv preprint arXiv:2305.02309}.

\bibitem[{Nijkamp et~al.(2022)Nijkamp, Pang, Hayashi, Tu, Wang, Zhou, Savarese, and Xiong}]{nijkamp2022codegen}
Erik Nijkamp, Bo~Pang, Hiroaki Hayashi, Lifu Tu, Huan Wang, Yingbo Zhou, Silvio Savarese, and Caiming Xiong. 2022.
\newblock Codegen: An open large language model for code with multi-turn program synthesis.
\newblock \emph{arXiv preprint arXiv:2203.13474}.

\bibitem[{OpenAI(2023{\natexlab{a}})}]{chatgpt:23}
OpenAI. 2023{\natexlab{a}}.
\newblock \href {https://openai.com/blog/chatgpt2} {Introducing chatgpt}.

\bibitem[{OpenAI(2023{\natexlab{b}})}]{openai2023gpt}
R~OpenAI. 2023{\natexlab{b}}.
\newblock Gpt-4 technical report. arxiv 2303.08774.
\newblock \emph{View in Article}, 2.

\bibitem[{Ouyang et~al.(2022)Ouyang, Wu, Jiang, Almeida, Wainwright, Mishkin, Zhang, Agarwal, Slama, Ray et~al.}]{ouyang2022training}
Long Ouyang, Jeffrey Wu, Xu~Jiang, Diogo Almeida, Carroll Wainwright, Pamela Mishkin, Chong Zhang, Sandhini Agarwal, Katarina Slama, Alex Ray, et~al. 2022.
\newblock Training language models to follow instructions with human feedback.
\newblock \emph{Advances in Neural Information Processing Systems}, 35:27730--27744.

\bibitem[{Patil et~al.(2023)Patil, Zhang, Wang, and Gonzalez}]{patil2023gorilla}
Shishir~G Patil, Tianjun Zhang, Xin Wang, and Joseph~E Gonzalez. 2023.
\newblock Gorilla: Large language model connected with massive apis.
\newblock \emph{arXiv preprint arXiv:2305.15334}.

\bibitem[{Pennington et~al.(2014)Pennington, Socher, and Manning}]{pennington2014glove}
Jeffrey Pennington, Richard Socher, and Christopher~D Manning. 2014.
\newblock Glove: Global vectors for word representation.
\newblock In \emph{Proceedings of the 2014 conference on empirical methods in natural language processing (EMNLP)}, pages 1532--1543.

\bibitem[{Qin et~al.(2023{\natexlab{a}})Qin, Hu, Lin, Chen, Ding, Cui, Zeng, Huang, Xiao, Han et~al.}]{qin2023tool}
Yujia Qin, Shengding Hu, Yankai Lin, Weize Chen, Ning Ding, Ganqu Cui, Zheni Zeng, Yufei Huang, Chaojun Xiao, Chi Han, et~al. 2023{\natexlab{a}}.
\newblock Tool learning with foundation models.
\newblock \emph{arXiv preprint arXiv:2304.08354}.

\bibitem[{Qin et~al.(2023{\natexlab{b}})Qin, Liang, Ye, Zhu, Yan, Lu, Lin, Cong, Tang, Qian, Zhao, Hong, Tian, Xie, Zhou, Gerstein, Li, Liu, and Sun}]{toolbench:23}
Yujia Qin, Shihao Liang, Yining Ye, Kunlun Zhu, Lan Yan, Yaxi Lu, Yankai Lin, Xin Cong, Xiangru Tang, Bill Qian, Sihan Zhao, Lauren Hong, Runchu Tian, Ruobing Xie, Jie Zhou, Mark Gerstein, Dahai Li, Zhiyuan Liu, and Maosong Sun. 2023{\natexlab{b}}.
\newblock \href {https://arxiv.org/abs/2307.16789} {Toolllm: Facilitating large language models to master 16000+ real-world apis}.

\bibitem[{Rae et~al.(2022)Rae, Borgeaud, Cai, Millican, Hoffmann, Song, Aslanides, Henderson, Ring, Young, Rutherford, Hennigan, Menick, Cassirer, Powell, van~den Driessche, Hendricks, Rauh, Huang, Glaese, Welbl, Dathathri, Huang, Uesato, Mellor, Higgins, Creswell, McAleese, Wu, Elsen, Jayakumar, Buchatskaya, Budden, Sutherland, Simonyan, Paganini, Sifre, Martens, Li, Kuncoro, Nematzadeh, Gribovskaya, Donato, Lazaridou, Mensch, Lespiau, Tsimpoukelli, Grigorev, Fritz, Sottiaux, Pajarskas, Pohlen, Gong, Toyama, de~Masson~d'Autume, Li, Terzi, Mikulik, Babuschkin, Clark, de~Las~Casas, Guy, Jones, Bradbury, Johnson, Hechtman, Weidinger, Gabriel, Isaac, Lockhart, Osindero, Rimell, Dyer, Vinyals, Ayoub, Stanway, Bennett, Hassabis, Kavukcuoglu, and Irving}]{rae2022scaling}
Jack~W. Rae, Sebastian Borgeaud, Trevor Cai, Katie Millican, Jordan Hoffmann, Francis Song, John Aslanides, Sarah Henderson, Roman Ring, Susannah Young, Eliza Rutherford, Tom Hennigan, Jacob Menick, Albin Cassirer, Richard Powell, George van~den Driessche, Lisa~Anne Hendricks, Maribeth Rauh, Po-Sen Huang, Amelia Glaese, Johannes Welbl, Sumanth Dathathri, Saffron Huang, Jonathan Uesato, John Mellor, Irina Higgins, Antonia Creswell, Nat McAleese, Amy Wu, Erich Elsen, Siddhant Jayakumar, Elena Buchatskaya, David Budden, Esme Sutherland, Karen Simonyan, Michela Paganini, Laurent Sifre, Lena Martens, Xiang~Lorraine Li, Adhiguna Kuncoro, Aida Nematzadeh, Elena Gribovskaya, Domenic Donato, Angeliki Lazaridou, Arthur Mensch, Jean-Baptiste Lespiau, Maria Tsimpoukelli, Nikolai Grigorev, Doug Fritz, Thibault Sottiaux, Mantas Pajarskas, Toby Pohlen, Zhitao Gong, Daniel Toyama, Cyprien de~Masson~d'Autume, Yujia Li, Tayfun Terzi, Vladimir Mikulik, Igor Babuschkin, Aidan Clark, Diego de~Las~Casas, Aurelia Guy, Chris Jones,
  James Bradbury, Matthew Johnson, Blake Hechtman, Laura Weidinger, Iason Gabriel, William Isaac, Ed~Lockhart, Simon Osindero, Laura Rimell, Chris Dyer, Oriol Vinyals, Kareem Ayoub, Jeff Stanway, Lorrayne Bennett, Demis Hassabis, Koray Kavukcuoglu, and Geoffrey Irving. 2022.
\newblock \href {http://arxiv.org/abs/2112.11446} {Scaling language models: Methods, analysis \& insights from training gopher}.

\bibitem[{Robertson et~al.(1995)Robertson, Walker, Jones, Hancock-Beaulieu, and Gatford}]{robertson1995okapi}
Stephen Robertson, S.~Walker, S.~Jones, M.~M. Hancock-Beaulieu, and M.~Gatford. 1995.
\newblock \href {https://www.microsoft.com/en-us/research/publication/okapi-at-trec-3/} {Okapi at trec-3}.
\newblock In \emph{Overview of the Third Text REtrieval Conference (TREC-3)}, pages 109--126. Gaithersburg, MD: NIST.

\bibitem[{Schick et~al.(2023)Schick, Dwivedi-Yu, Dessì, Raileanu, Lomeli, Zettlemoyer, Cancedda, and Scialom}]{schick2023toolformer}
Timo Schick, Jane Dwivedi-Yu, Roberto Dessì, Roberta Raileanu, Maria Lomeli, Luke Zettlemoyer, Nicola Cancedda, and Thomas Scialom. 2023.
\newblock \href {http://arxiv.org/abs/2302.04761} {Toolformer: Language models can teach themselves to use tools}.

\bibitem[{Shen et~al.(2023{\natexlab{a}})Shen, Song, Tan, Li, Lu, and Zhuang}]{shen2023hugginggpt}
Yongliang Shen, Kaitao Song, Xu~Tan, Dongsheng Li, Weiming Lu, and Yueting Zhuang. 2023{\natexlab{a}}.
\newblock Hugginggpt: Solving ai tasks with chatgpt and its friends in huggingface.
\newblock \emph{arXiv preprint arXiv:2303.17580}.

\bibitem[{Shen et~al.(2023{\natexlab{b}})Shen, Song, Tan, Zhang, Ren, Yuan, Lu, Li, and Zhuang}]{shen2023taskbench}
Yongliang Shen, Kaitao Song, Xu~Tan, Wenqi Zhang, Kan Ren, Siyu Yuan, Weiming Lu, Dongsheng Li, and Yueting Zhuang. 2023{\natexlab{b}}.
\newblock Taskbench: Benchmarking large language models for task automation.
\newblock \emph{arXiv preprint arXiv:2311.18760}.

\bibitem[{Singh et~al.(2022)Singh, Blukis, Mousavian, Goyal, Xu, Tremblay, Fox, Thomason, and Garg}]{singh2022progprompt}
Ishika Singh, Valts Blukis, Arsalan Mousavian, Ankit Goyal, Danfei Xu, Jonathan Tremblay, Dieter Fox, Jesse Thomason, and Animesh Garg. 2022.
\newblock \href {http://arxiv.org/abs/2209.11302} {Progprompt: Generating situated robot task plans using large language models}.

\bibitem[{Taori et~al.(2023)Taori, Gulrajani, Zhang, Dubois, Li, Guestrin, Liang, and Hashimoto}]{alpaca}
Rohan Taori, Ishaan Gulrajani, Tianyi Zhang, Yann Dubois, Xuechen Li, Carlos Guestrin, Percy Liang, and Tatsunori~B. Hashimoto. 2023.
\newblock Stanford alpaca: An instruction-following llama model.
\newblock \url{https://github.com/tatsu-lab/stanford_alpaca}.

\bibitem[{Thoppilan et~al.(2022)Thoppilan, De~Freitas, Hall, Shazeer, Kulshreshtha, Cheng, Jin, Bos, Baker, Du et~al.}]{thoppilan2022lamda}
Romal Thoppilan, Daniel De~Freitas, Jamie Hall, Noam Shazeer, Apoorv Kulshreshtha, Heng-Tze Cheng, Alicia Jin, Taylor Bos, Leslie Baker, Yu~Du, et~al. 2022.
\newblock Lamda: Language models for dialog applications.
\newblock \emph{arXiv preprint arXiv:2201.08239}.

\bibitem[{Touvron et~al.(2023)Touvron, Lavril, Izacard, Martinet, Lachaux, Lacroix, Rozi{\`e}re, Goyal, Hambro, Azhar et~al.}]{touvron2023llama}
Hugo Touvron, Thibaut Lavril, Gautier Izacard, Xavier Martinet, Marie-Anne Lachaux, Timoth{\'e}e Lacroix, Baptiste Rozi{\`e}re, Naman Goyal, Eric Hambro, Faisal Azhar, et~al. 2023.
\newblock Llama: Open and efficient foundation language models.
\newblock \emph{arXiv preprint arXiv:2302.13971}.

\bibitem[{Valmeekam et~al.(2023)Valmeekam, Sreedharan, Marquez, Olmo, and Kambhampati}]{valmeekam2023planning}
Karthik Valmeekam, Sarath Sreedharan, Matthew Marquez, Alberto Olmo, and Subbarao Kambhampati. 2023.
\newblock On the planning abilities of large language models (a critical investigation with a proposed benchmark).
\newblock \emph{arXiv preprint arXiv:2302.06706}.

\bibitem[{Wang et~al.(2017)Wang, Mao, Wang, and Guo}]{wang8047276}
Quan Wang, Zhendong Mao, Bin Wang, and Li~Guo. 2017.
\newblock \href {https://doi.org/10.1109/TKDE.2017.2754499} {Knowledge graph embedding: A survey of approaches and applications}.
\newblock \emph{IEEE Transactions on Knowledge and Data Engineering}, 29(12):2724--2743.

\bibitem[{Wang et~al.(2022)Wang, Wei, Schuurmans, Le, Chi, Narang, Chowdhery, and Zhou}]{wang2022self}
Xuezhi Wang, Jason Wei, Dale Schuurmans, Quoc Le, Ed~Chi, Sharan Narang, Aakanksha Chowdhery, and Denny Zhou. 2022.
\newblock Self-consistency improves chain of thought reasoning in language models.
\newblock \emph{arXiv preprint arXiv:2203.11171}.

\bibitem[{Wei et~al.(2022{\natexlab{a}})Wei, Tay, Bommasani, Raffel, Zoph, Borgeaud, Yogatama, Bosma, Zhou, Metzler et~al.}]{wei2022emergent}
Jason Wei, Yi~Tay, Rishi Bommasani, Colin Raffel, Barret Zoph, Sebastian Borgeaud, Dani Yogatama, Maarten Bosma, Denny Zhou, Donald Metzler, et~al. 2022{\natexlab{a}}.
\newblock Emergent abilities of large language models.
\newblock \emph{arXiv preprint arXiv:2206.07682}.

\bibitem[{Wei et~al.(2022{\natexlab{b}})Wei, Wang, Schuurmans, Bosma, Xia, Chi, Le, Zhou et~al.}]{wei2022chain}
Jason Wei, Xuezhi Wang, Dale Schuurmans, Maarten Bosma, Fei Xia, Ed~Chi, Quoc~V Le, Denny Zhou, et~al. 2022{\natexlab{b}}.
\newblock Chain-of-thought prompting elicits reasoning in large language models.
\newblock \emph{Advances in Neural Information Processing Systems}, 35:24824--24837.

\bibitem[{Wieting et~al.(2015)Wieting, Bansal, Gimpel, and Livescu}]{wieting2015towards}
John Wieting, Mohit Bansal, Kevin Gimpel, and Karen Livescu. 2015.
\newblock Towards universal paraphrastic sentence embeddings.
\newblock \emph{arXiv preprint arXiv:1511.08198}.

\bibitem[{Wolf et~al.(2019)Wolf, Debut, Sanh, Chaumond, Delangue, Moi, Cistac, Rault, Louf, Funtowicz, Davison, Shleifer, von Platen, Ma, Jernite, Plu, Xu, Scao, Gugger, Drame, Lhoest, and Rush}]{huggingfacetransformers}
Thomas Wolf, Lysandre Debut, Victor Sanh, Julien Chaumond, Clement Delangue, Anthony Moi, Pierric Cistac, Tim Rault, Rémi Louf, Morgan Funtowicz, Joe Davison, Sam Shleifer, Patrick von Platen, Clara Ma, Yacine Jernite, Julien Plu, Canwen Xu, Teven~Le Scao, Sylvain Gugger, Mariama Drame, Quentin Lhoest, and Alexander~M. Rush. 2019.
\newblock \href {https://doi.org/10.48550/ARXIV.1910.03771} {Huggingface's transformers: State-of-the-art natural language processing}.

\bibitem[{Wu et~al.(2023)Wu, Min, Bisk, Salakhutdinov, Azaria, Li, Mitchell, and Prabhumoye}]{pet:23}
Yue Wu, So~Yeon Min, Yonatan Bisk, Ruslan Salakhutdinov, Amos Azaria, Yuanzhi Li, Tom Mitchell, and Shrimai Prabhumoye. 2023.
\newblock \href {https://arxiv.org/abs/2305.02412} {Plan, eliminate, and track — language models are good teachers for embodied agents.}

\bibitem[{Xie et~al.(2021)Xie, Raghunathan, Liang, and Ma}]{xie2021explanation}
Sang~Michael Xie, Aditi Raghunathan, Percy Liang, and Tengyu Ma. 2021.
\newblock An explanation of in-context learning as implicit bayesian inference.
\newblock \emph{arXiv preprint arXiv:2111.02080}.

\bibitem[{Yao et~al.(2022)Yao, Zhao, Yu, Du, Shafran, Narasimhan, and Cao}]{yao2022react}
Shunyu Yao, Jeffrey Zhao, Dian Yu, Nan Du, Izhak Shafran, Karthik Narasimhan, and Yuan Cao. 2022.
\newblock React: Synergizing reasoning and acting in language models.
\newblock \emph{arXiv preprint arXiv:2210.03629}.

\bibitem[{Zeng et~al.(2022)Zeng, Liu, Du, Wang, Lai, Ding, Yang, Xu, Zheng, Xia et~al.}]{zeng2022glm}
Aohan Zeng, Xiao Liu, Zhengxiao Du, Zihan Wang, Hanyu Lai, Ming Ding, Zhuoyi Yang, Yifan Xu, Wendi Zheng, Xiao Xia, et~al. 2022.
\newblock Glm-130b: An open bilingual pre-trained model.
\newblock \emph{arXiv preprint arXiv:2210.02414}.

\bibitem[{Zhang et~al.(2022)Zhang, Roller, Goyal, Artetxe, Chen, Chen, Dewan, Diab, Li, Lin et~al.}]{zhang2022opt}
Susan Zhang, Stephen Roller, Naman Goyal, Mikel Artetxe, Moya Chen, Shuohui Chen, Christopher Dewan, Mona Diab, Xian Li, Xi~Victoria Lin, et~al. 2022.
\newblock Opt: Open pre-trained transformer language models.
\newblock \emph{arXiv preprint arXiv:2205.01068}.

\bibitem[{Zhang et~al.(2016)Zhang, Rahman, Braylan, Dang, Chang, Kim, McNamara, Angert, Banner, Khetan et~al.}]{zhang2016neural}
Ye~Zhang, Md~Mustafizur Rahman, Alex Braylan, Brandon Dang, Heng-Lu Chang, Henna Kim, Quinten McNamara, Aaron Angert, Edward Banner, Vivek Khetan, et~al. 2016.
\newblock Neural information retrieval: A literature review.
\newblock \emph{arXiv preprint arXiv:1611.06792}.

\end{thebibliography}
